\documentclass[aps,pra,showpacs,superscriptaddress,floatfix]{revtex4}
\usepackage{epsfig}\usepackage{amsmath}
\usepackage{amssymb}

\newcommand{\id}{1}
\newcommand{\identity}{\openone}
\newcommand{\be}{\begin{equation}}
\newcommand{\bea}{\begin{eqnarray}}
\newcommand{\eea}{\end{eqnarray}}
\newcommand{\ee}{\end{equation}}

\newcommand{\bra}[1]{\mbox{$\langle #1 |$}}
\newcommand{\ket}[1]{\mbox{$| #1 \rangle$}}

\begin{document}

\title{Remote Implementation of Quantum Operations}

\author {Susana F. Huelga}
\affiliation{Quantum Physics Group, STRI,Department of Physics,
Astronomy and Mathematics, University of Hertfordshire, Hatfield,
Herts AL10 9AB, UK}
\author{Martin B. Plenio}
\affiliation{QOLS, Blackett Laboratory, Imperial College London,
London SW7 2BW, UK} \affiliation{Institute for Mathematical
Sciences, Imperial College London, 53 Exhibition Road, London SW7
2BW, UK}
\author{Guo-Yong Xiang}\author{Jian Li} \author{Guang-Can Guo}
\affiliation{Key Laboratory of Quantum Information and Department
of Physics,\\
University of Science and Technology of China, Hefei 230026,
China}
%\maketitle
\date{\today}

\begin{abstract}
Shared entanglement allows, under certain conditions, the remote
implementation of quantum operations. We revise and extend recent
theoretical results on the remote control of quantum systems as
well as experimental results on the remote manipulation of
photonic qubits via linear optical elements.

\end{abstract}
 \maketitle

\section{Introduction}
Consider a maximally entangled state of two particles $A$ and $B$,
for instance, the Bell state
\begin{equation}
\ket{\Phi^+}_{AB}=\frac{1}{\sqrt{2}} (\ket{0}_A\ket{0}_B +
\ket{1}_A\ket{1}_B). \label{bell}
\end{equation}
Imagine now that the state describes a situation where the
particles are in space-like separated regions and let us call
Alice and Bob the remote partners with access to particles $A$ and
$B$ respectively. We assume that Alice and Bob can perform
arbitrary local operations (LO) on their subsystems, including
unitary actions and, possibly, generalized measurements involving
extra ancilla particles. If necessary, Alice and Bob can exchange
classical communication (CC) but global quantum operations
involving subsystems $A$ and $B$ are forbidden. The resulting set
of allowed operations is generally referred to as LOCC. When Alice
and Bob share entanglement, local actions on a given region have a
{\em non-local} effect, in the sense that the state of the remote
particle does not necessarily remain insensitive to the details of
the transformation performed miles away. Consider the case of a
projective measurement on Alice's side. For example, she performs
a Stern-Gerlach measurement with the apparatus oriented along an
arbitrary unit vector $\hat{n}$. Let us call $\ket{+}_{\hat{n}}$,
the eigenvector corresponding to spin up along the $\hat{n}$
direction. Whenever she records a spin up result, the joint state
of the system is projected out into the product state
$\ket{+}_{\hat{n},A} \ket{+}_{\hat{n},B}$, if she records a spin
down result, the joint state of the system is projected out into
the product state $\ket{-}_{\hat{n},A} \ket{-}_{\hat{n},B}$. Both
outcomes happen with equal probability.
\begin{figure}[htb]
\epsfxsize=6.7cm
\begin{center}
\epsffile{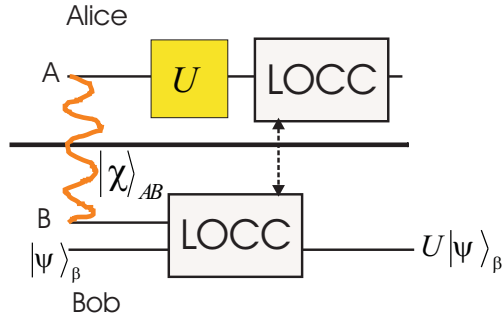}
\end{center}
\caption{\label{effprob} Two remote partners share bipartite
entanglement, represented by the state $\ket{\chi}_{AB}$. Only one
of the partners, Alice, has access to a device that can implement
an arbitrary single qubit operation $U$. Is there an LOCC protocol
that can yield a final state where Bob ends up holding the state
$U\ket{\psi}$ for any $\ket{\psi}$? }
\end{figure}
Now there are two distinctly different ways to proceed. If Bob
were to conduct an independent measurement on his particles using
an Stern-Gerlach apparatus with a random orientation, then he will
also register a random outcome, obtaining an unbiased spin up and
spin down distribution along his chosen direction. There will be
no correlations between his measurement outcomes and those of
Alice. If however, Bob receives a message from Alice informing him
of her choice of alignment, then the situation changes in a subtle
way. His measurement will still give random outcomes, but, these
outcomes will be perfectly correlated to those of Alice's
measurement outcomes. This shows that classical communication may
be used to correlate actions between partners with the effect that
experimental results will change significantly. In a more
complicated setting Alice and Bob would perform measurements along
various non-parallel but correlated axes. This type of correlated
measurements then leads to the celebrated EPR paradox, later
formulated by Bell in a testable form and which for years fuelled
the discussions about quantum non-locality \cite{peres}. The
recent development of the theory of quantum information processing
has given an unexpected twist to this discussion. Shared
entanglement supplement by LOCC operations makes possible entirely
new forms of distributed computation and communication
\cite{review}, as epitomized by the process of quantum state
teleportation \cite{tele}. In this article we will focus on a
related problem: The remote implementation of quantum operations,
and in particular, the teleportation of unitary transformations
\cite{teleU,angles}. The simplest situation has been illustrated
pictorially in Figure 1.
One of the remote partners, Alice, has access to a classical
device that can implement an arbitrary singe-particle operation
$U$. There is no need to restrict the dimensionality of the space
of allowed operations except for assuming that it be finite, but
possibly very large (See section I for specific details). Using
LOCC operations, and provided that Alice and Bob share
entanglement, represented by the state $\ket{\chi}_{\alpha AB}$ in
the figure, our aim is to remotely implement the arbitrary
transformation $U$ on Bob's side, i.e., to design an LOCC protocol
that will end up with Bob holding the state $U \ket{\psi}_{\beta}$
for any single particle state $\ket{\psi}_{\beta}$. When we allow
for a completely arbitrary $U$ then we will show that the most
economical procedure to achieve this consist of teleporting the
state of Bob's particle to Alice, who applies the transformation
$U$ to the teleported state and then teleports the state back to
Bob. No other LOCC protocol supplemented by shared entanglement
will consume less distributed entanglement and have a lower
communication cost. However, when the requirement of perfect
remote control for arbitrary operations is relaxed, we encounter a
different situation and state teleportation is no longer the most
cost-efficient way to proceed when implementing restricted sets of
operations. Non-trivial protocols become possible and we present
examples for these in the special case of unitary transformations
on two-level systems (qubits).

We have organized the presentation as follows. In Section II we
present a general proof for the necessary and sufficient resources
required for the remote control of a single qudit operation. This
results generalize previous constraints derived for qubits
\cite{teleU}. Section III analyzes the limitations arising when
one attempts to teleport restricted sets of operations. We will
show that arbitrary qubit rotations around a fixed direction can
be implemented remotely without the need of teleporting states
between Alice and Bob. Novel results concerning the remote control
of identical copies are discussed in III.a. Recent experimental
reports demonstrating the protocol are summarized in section VI.
The final section summarizes our main results and conclusions.

%%%%%%%%%%%%%%%%%%%%%%%%%%%%%%%%%%%%%%%%%%%%%%%%%%%%%%%%%%%%%%%%%%%%%%%%%%%%%%%%%% NO REMOTE %%%%%%%%%%%%%%%%%%%%%%%%%%%%%%%%%%%
\section{A general result}
The most general scenario for the teleportation of an arbitrary
unitary operation was discussed in \cite{teleU} and is related to
the design of universal quantum gates arrays proposed by Nielsen
and Chuang \cite{nielsen}. When the device implementing the
unitary transformation $U$ is modelled as a truly quantum system,
whose state corresponding to the unitary operation {\em U} will be
denoted by $\ket{U}_{C}$, we can represent the remote control
operator by a completely positive, linear, trace preserving map on
the set of density operators for the combined system. Any such map
has a unitary representation $\cal G$ involving global ancillary
systems in state $\ket{\chi}_{AB}$, so that
\begin{equation}
    {\cal G} \big[\ket{\chi}_{AB}{\otimes}\ket{U}_{C}{\otimes}\ket{\psi}_{\beta}\big]=
    \ket{\Phi(U,{\chi})}_{ABC}{\otimes}\left(U\ket{\psi}_{\beta}\right)\ ..
    \label{tony}
\end{equation}
Note that this unitary representation may be non-local even if the
map itself is local. As a consequence, any argument involving the
eq. (\ref{tony}) will provide only lower bounds on the resource
requirement. In any specific case one will then need to find a
local protocol that matches these lower bounds. This can indeed be
done in any of the instances that we are considering in the
following.
\begin{figure}[htb]
\epsfxsize=8.7cm
\begin{center}
\epsffile{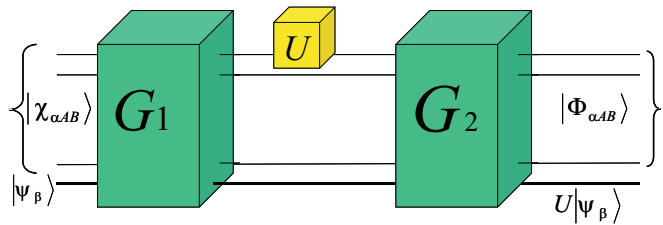}
\end{center}
\caption{\label{circuito} Quantum circuit model using nonlocal
unitary gates $G_1$ and $G_2$ for the teleportation of a unitary
operation $U$ on a single particle (See text for details). Alice
and Bob have access to the two upper and the two lower wires
respectively. The lowest wire represents the selected single
particle state onto which we intend to remotely apply the
transformation $U$.}
\end{figure}
The results by Nielsen and Chuang on programmable gate arrays
\cite{nielsen} imply that the control states $\ket{U}_C$
corresponding to different unitary transformations are orthogonal,
so that no finite-dimensional control system can be used to
teleport an arbitrary unitary operation. For the remainder of this
paper, when we speak of an arbitrary unitary operation, we will
mean one which belongs to some arbitrarily large, but finite, set.
We will also assume that this set contains the identity
$\sigma^0={\identity}$ and the 3 Pauli operators ${\sigma}^{i}$.
The orthogonality of the control states opens the possibility that
different operations can, at least in principle, be distinguished
and identified by Alice if she chooses to perform measurements on
the apparatus. Here we are not interested in this approach and
focused instead in assisting the task with shared entangled and
LOCC operations \cite{bcn}.

Within this very general formulation, it is possible to derive
lower bounds on the amount of non-local resources that are needed
to implement ${\cal G}$ using only local operations and classical
communication. To this end, we should remember the following basic
principles \cite{poli,teleU}\\

\noindent{\bf (a)} {\em The amount of classical information able
to be communicated by an operation in a given direction across
some partition between subsystems cannot exceed the amount of
information that must be sent in this direction across the same
partition to complete the operation} (Impossibility of
super-luminal communication).\\

\noindent {\bf (b)} {\em The amount of bipartite entanglement that
an operation can establish across some partition between
subsystems cannot exceed the amount of prior entanglement across
the partition that must be consumed in order to complete the
operation}(Impossibility of increasing entanglement under LOCC).\\

Principle (a) allows us to establish the fact that at least two
classical bits must be sent from Alice to Bob to complete the
teleportation of an arbitrary {\em U}. Moreover, by teleporting an
arbitrary {\em U} according to the general prescription in Eq.
(\ref{tony}), Alice and Bob can establish 2 ebits of shared
entanglement and it follows from principle (b) that at least 2
ebits of entanglement need to be consumed to implement ${\cal G}$
locally, i.e. to teleport an arbitrary unitary operation
\cite{teleU,angles}. These bounds can be attained by a procedure
in which Bob teleports the state of his particle to Alice who,
after applying the unitary transformation, teleports it back to
him ( bidirectional state teleportation). This scheme saturates
the lower bounds for the amount of shared ebits and classical bits
transmitted from Alice to Bob and additionally uses two bits of
classical communication from Bob to Alice and allows the faithful
implementation of $U$ independently of the dimension of the
control system.  To be more efficient overall, any other scheme
would need less resources than bidirectional state teleportation.
This establishes an upper bound in the overall amount of resources
required for the efficient remote implementation of an arbitrary
$U$ as 4 classical bits and 2 ebits.

Note that it would also be conceivable to adopt a different
strategy -- teleporting the state of the control system from Alice
to Bob who would then implement the control directly onto qubit
$\beta$. We call this the ``control-state teleportation'' scheme.
Control-state teleportation is a unidirectional communication
scheme from Alice to Bob, so the absolute lower bound for the
communication exchange from Bob to Alice is zero. Obviously, the
overall resources will depend on the dimensionality of the control
system $C$ and in general a large amount of entanglement and
classical communication from Alice to Bob will be required if we
want to teleport the control system.

Let us focus on the experimental scenario where the black box
implementing an arbitrary transformation $U$ is a macroscopic
object, involving a (very) large number of degrees of freedom. The
option of teleporting the control apparatus is then unfeasible,
given that it would consume an unrealistic amount of entanglement
and classical communication resources. However, the question
remains whether there exists a more economical protocol than
bidirectional state teleportation. We will generalize in the
following the results presented in \cite{teleU} and show that
bidirectional state teleportation is an unconditional optimal way
to remotely implement an arbitrary $U$ on a given $d$-level system
(qudit).

Discarding the possibility of control-state teleportation allows
us to replace the transformation given by Eq. (\ref{tony}) with
%******************************************************************
\begin{equation}
  G_2 \, U \, G_1 (\ket{\chi}_{\alpha AB} \otimes\ket{\psi}_{\beta})
   =  \ket{\Phi(U,\chi)}_{\alpha AB} \otimes U \ket{\psi}_{\beta}
  \label{doble},
\end{equation}
where certain fixed operations $G_1$ and $G_2$ are performed,
respectively, prior to and following the action of the arbitrary
$U$ on a qubit $\alpha$ on Alice's side, as illustrated in Figure
2. We assume that Alice and Bob share initially some entanglement,
represented by the state $\ket{\chi}_{\alpha AB}$. As before, the
purpose of the transformation is to perform the operation $U$ on
Bob's qubit $\beta$. Note that we chose to use a nonlocal unitary
representation of the transformation involved so that $G_1$ and
$G_2$ are unitary operators acting on possibly all subsystems. For
instance, the transformation $G_i$ can represent a state
teleportation operation.
%%%%%%%%%%%%%%%%%%%%%%%%%%%%%%%%%%%%%%%%%%%%%%%%%%%%%%%%%%%%%%%%%%%%%%%%%%%%%%%%
%%%%%%%%%%%%%%%%%%%%%%%%%%%%%%%%%%%%%%%%%%%%%%%%%%%%%%%%%%%%%%%%%%%%%%%%%%%%%%%%%%%%%
In the following we prove, for systems of arbitrary spatial
dimensions, that this is necessarily the case and that the only
way that Eq. (\ref{doble}) can be implemented (locally) is by
teleporting the state $\ket{\psi}_\beta$ from Bob to Alice, and
then teleporting back the transformed state $U\ket{\psi}_\beta$
from Alice to Bob.

By linearity, the transformed state of systems $\alpha AB$ has to
be independent of the particular input state $\ket{\psi}_{\beta}$.
and the specific operation {\em U} \cite{teleU}.  With this we can
already show that the operation $G_1$ cannot be trivial. We do
this by first assuming the contrary that $G_1=\identity$, and
considering two input states, $\ket{\psi}_{\beta}$ and
$\ket{\psi'}_{\beta}$ such that
$_{\beta}{\langle}{\psi}^{'}|{\psi}{\rangle}_{\beta}=0$, and two
unitary transformations $U$ and $U'$ which bring these two states
to the same state $\ket{\gamma}_{\beta}$.  Using Eq.
(\ref{doble}), this implies that
\begin{eqnarray}
  G_2 \left( U\ket{\chi}_{\alpha AB} \, \ket{\psi}_{\beta} \right)
      &=& \ket{\Phi(\chi)}_{\alpha AB}{\otimes}\ket{\gamma}_{\beta} \nonumber \\
  G_2 \left( U' \ket{\chi}_{\alpha AB} \, \ket{\psi'}_{\beta} \right)
      &=& \ket{\Phi(\chi)}_{\alpha AB}{\otimes} \ket{\gamma}_{\beta}\ ..
  \label{chu}
\end{eqnarray}
No universal unitary action $G_2$ can be found to satisfy Eq.
(\ref{chu}), as this would require the mapping of orthogonal
states onto the same state. This shows that no universal operation
$G_2$ that satisfies Eq. (\ref{chu}) can exist and therefore, for
the $U$-teleportation to succeed, $G_1$ has to be non-trivial.
Finally, we will show that each of the operations $G_i$ implements
a state transfer. Let us rewrite Eq. (\ref{doble}) as
\begin{equation}
U \, G_1 (\ket{\chi}_{\alpha AB} \otimes \ket{\psi}_{\beta}) =
G^\dagger_2(\ket{\Phi(\chi)}_{\alpha AB} \otimes U
\ket{\psi}_{\beta}). \label{doble2}
\end{equation}
%%%%%%%%%%%%%%%%%%%%%%%%%%%%%%%%%%%%%%%%%%%%%%%%%%%%%%%%%%%%%%%%%%%%%%%% Extension to qudits %%%%%%%%%%%%%%%%%%%%%%%%%%%%%%%%%%%%%%%%%%********&&&&&&^^^^
Let us denote by $\{ \ket{k} \}_{k=0}^{d-1}$ the canonical basis
in the space of sates of qudits. Consider the hermitean operator
$\Pi$ defined as
\begin{equation}
\Pi=\sum_{k=0}^{d-1} (k+1) \ket{k} \bra{k}.
\end{equation}
By construction $\Pi$ is diagonal in the canonical basis and has a
non-degenerate spectrum. Since the operations $G_1$ and $G_2$ are
universal, we may choose $U$ and $\ket{\psi}_\beta$ freely. For
each $\ket{\psi}_\beta$ let the operator $U_\psi$ be such that
$U_\psi\ket{\psi}=\ket{0}$ where $\Pi \ket{0}=\ket{0}$. If $U= \Pi
U_{\psi}$, then
\begin{eqnarray*}
  \left(\Pi U_\psi\right) G_1\! \left(\ket{\chi}_{\alpha AB}
              \!\otimes\ket{\psi}_\beta\right)
  \!\!&=&\! G^\dagger_2\!\left(\ket{\Phi(\chi)}_{\alpha AB}\!\otimes
              \Pi U_\psi \ket{\psi}_\beta \right)\\
  \!\!&=&\! G^\dagger_2\!\left(\ket{\Phi(\chi)}_{\alpha AB}\otimes\ket{0}_\beta\right)\ .
\end{eqnarray*}
The RHS is simply $(U_\psi) G_1\left(\ket{\chi}_{\alpha
AB}\otimes\ket{\psi}_\beta\right)$ and so, given that the spectrum
of the operator $\Pi$ is not degenerate, $(U_\psi)
G_1\left(\ket{\chi}_{\alpha AB}\otimes\ket{\psi}_\beta\right)$ is
the eigenstate $\ket{0}_{\alpha} \otimes\ket{\phi}_{AB \beta}$ of
$\left(\Pi \right)_{\alpha} \otimes\id_{AB \beta}$. Equivalently,
\begin{eqnarray}
  G_1\left(\ket{\chi}_{\alpha AB}\otimes\ket{\psi}_\beta\right)
  &=& \left(U^\dagger_\psi\ket{0}_\alpha\right)\otimes\ket{\phi}_{AB \beta} \nonumber\\
  &=& \ket{\psi}_\alpha\otimes\ket{\phi}_{AB \beta} \label{tele}\ .
\end{eqnarray}
%&&&&&&&&&&&&&&&&&&&&&&&&&&&& END %%%%%%%%%%%%%%%%%%%%%%%%%%%%%%%%%%%%%%%%%%%%%%%%%%%%%%%%%%%%%%%%%%%%%qudits%%%%%%%%%%%%%%%%%%%%%%%%%%%%%%%%%%%%%%
In other words, the operation $G_1$ necessarily transfers Bob's
state $\ket{\psi}$ to Alice's qubit $\alpha$. Substituting Eq.
(\ref{tele}) into Eq. (\ref{doble}) then shows that $G_2$
necessarily transfers $U\ket{\psi}$ back to Bob's qubit ${\beta}$.
This implies that the state of Bob's qubit must be brought to
Alice for it to be acted on by the local operator $U$. This
results is tantamount to a no-go theorem: {\it a local unitary
operation $U$ cannot act remotely.} From this and the fact that
quantum state teleportation is an optimal procedure for local
state transfer, we conclude that the optimal LOCC procedure
supplemented by shared entanglement for implementing remotely an
arbitrary unitary action $U$ on a qudit is by means of
bidirectional state teleportation.

%%%%%%%%%%%%%%%%%%%%&&&&&&&&&&&&&&&&&&&&&&&&&&&&&&&&&&&&&&&&&&&&&&&&&&&&&&&&&&&&&&&&&&&& RESTRICTED OPERATIONS $$$$$$$$$$$$$$$$$$$$$$$$$$$$$$$$$$$$$$$$$$$$
\section{Restricted operations on qubits: Teleporting arbitrary rotations around a fixed axis}

In the specific case of qubits, the results presented in the
previous section show that if we want the transformation $U$ to be
an arbitrary element of the group $SU(2)$, no LOCC protocol can
exist consuming less overall resources than teleporting Bob's
state to Alice followed by Alice teleporting the state transformed
by $U$ back to Bob. This amounts to two e-bits of entanglement and
two classical bit in each direction. The ultimate reason for this
result can be found in the linearity of quantum mechanics and the
impossibility of implementing remotely an arbitrary $U$ without
resorting to state transfer belongs to the family of no-go results
imposed by the linear structure of quantum mechanics
such as the non-cloning theorem \cite{cloning}.\\
The same way that the no-cloning theorem forbids the replication
of general states, should we expect a similar result if the
requirement of being able to implement {\em any} $U$ is relaxed?
Can we find families of operators that can be implemented
consuming less overall resources than a two-way teleportation
protocol?. We want the procedure to work with perfect efficiency.
Imperfect storage of quantum operations have been recently
discussed by Vidal et al. \cite{guifre}. The probabilistic
implementation of universal quantum processors was discussed by
Nielsen and Chuang \cite{nielsen} and more recently by Hillery et
al in the general case of qudits \cite{buzek}. We will show that
there are indeed two restricted classes of operations that can be
implemented remotely and deterministically using less overall
resources than bidirectional quantum state teleportation and only
two (up to a local change of basis). These are arbitrary rotations
around a fixed direction $\vec{n}$ and rotations by a fixed angle
around an arbitrary direction lying in a plane orthogonal to
$\vec{n}$ \cite{angles}. It is easy to show that if a given
protocol would be able to implement remotely certain operation
$U$, the same protocol would also allow the
implementation of a remote control-$U$. %%%%%%%%%%%%%%%%%%%%%%%%%%%%%%%%%%%%%%%%%%%%%%% CNOT %%%%%%%%%%%%%%%%%%%%%%%%%%%%%%%%%%%%%%%%
This argument will allow us to establish a lower bound on the
amount of classical communication that needs to be conveyed from
Bob to Alice. Consider the case of a non-local controlled-NOT gate
between Alice and Bob \cite{poli,sandu}. When Alice prepares the
state $\ket{+}_c=(\ket{0}+\ket{1})/\sqrt{2}$, the action of a
controlled-NOT gate with Bob qubit being in either state
$\ket{+}_B$ or in state $\ket{-}_B$ is given by
\begin{eqnarray}
\ket{+}_c \ket{+}_B &\longmapsto& \ket{+}_c \ket{+}_B, \\
\nonumber \ket{+}_c \ket{-}_B &\longmapsto& \ket{-}_c \ket{-}_B .
\end{eqnarray}
Therefore, this operation allows Bob to transmit one bit of
information to Alice and, as a consequence, the teleportation of
$U$ requires at least one bit of communication from Bob to Alice.
Summarizing, the physical principles of non-increase of
entanglement under LOCC and the impossibility of super-luminal
communication allow us to establish lower bounds in the resources
required for teleporting an unknown quantum operation on a qubit.
At least two e-bits of entanglement have to be consumed and, in
addition, this quantum channel has to be supplemented by a {\em
two way} classical communication channel which, in principle,
could be non-symmetric. While consistency with causality requires
two classical bits being transmitted from Alice to Bob, the lower
bound for the amount of classical information transmitted from Bob
to Alice has been found to be one bit.\\
An explicit protocol saturating these bounds was presented in
\cite{angles}. This protocol allows the remote implementation of
arbitrary rotations around a given axis $\hat{n}$ as well as fixed
rotations around an arbitrary direction within a plane orthogonal
to $\hat{n}$. Choosing $\hat{n}$ to be along the $z$-direction,
operations of the form
\begin{equation}
    U_{com} = \left(
    \begin{array}{cc}
    a & 0\\
    0 &  a^*
    \end{array}
    \right) = e^{i\frac{\phi}{2}\sigma_z},
    \label{uu1}
\end{equation}
with $a=e^{i\phi}$ that is, the set of operations that commute
with $\sigma_z$, or transformations of the form
\begin{equation}
    U_{anticom} = \left(
    \begin{array}{cc}
    0 & b\\
    -b^* &  0
    \end{array}
    \right)=\sigma_x e^{i(\frac{\phi + \pi}{2})\sigma_z} {\rm ,}
    \label{uu2}
\end{equation}
which anticommute with $\sigma_z$, i.e., are linear combinations
of the Pauli operators $\sigma_x$ and $\sigma_y$. Any operation
within this family can be teleported deterministically using a
protocol which employs less resources than bidirectional state
teleportation. Remarkably, these are the only sets that have this
property as we rigourously showed in \cite{angles}.
\begin{figure}[th]
\vspace*{0.5cm}
\centerline{\includegraphics[width=8cm]{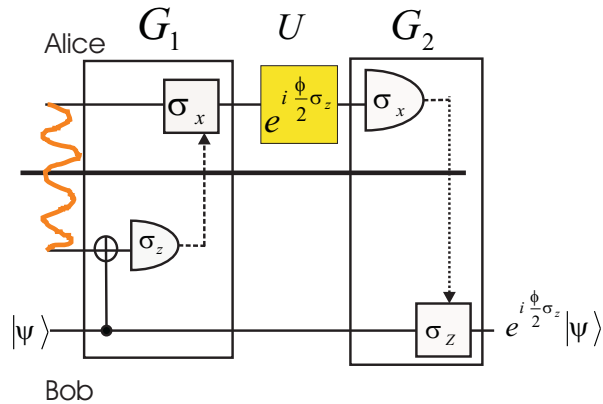} }
\vspace{.5cm} \caption{\label{gates} Quantum circuit for the
remote rotation of a single qubit. The whole process is divided
into three steps whose experimental implementation using photonic
qubits will be described in section IV. The operation $G_1$ makes
the coefficients of Bob's state visible in the channel via a
global action on Bob's side and subsequent measurement in the
computational basis. One classical bit has to be conveyed from
Alice to Bob. For instance, if Alice and Bob share initially the
Bell state $\ket{\phi}_{AB}^+$ and $\ket{\psi}=\alpha
\ket{0}_B+\beta \ket{1}_B$, after completing $G_1$ they share the
state $\alpha \ket{00}_{AB}+\beta \ket{11}_B$. Next, Alice
implements the rotation $U$ on her side. Operation $G_2$ involves
a measurement in the rotated basis followed by the transmission of
a classical bit to Bob, who completes the protocol by applying a
final $\sigma_z$ operation conditional on Alice's measurement
result. }
\end{figure}
The entanglement and communication costs can be further reduced if
it is a priory known whether the operation $U$ to be teleported
belongs to either the set characterized by eq. (\ref{uu1}) or to
the set characterized by eq. (\ref{uu2}). Figure 3 depicts the
quantum circuit representation of the protocol that allows the
deterministic remote implementation of arbitrary rotations around
the $z$-axis. This is a communication symmetrical protocol where
one classical bit is conveyed in each direction and with an
overall entanglement cost of 1 ebit. As always, wiggly lines
represent shared entanglement across the space-like thick solid
line. The dotted arrow lines represent the exchange of classical
communication following the measurement of the observable enclosed
in the half-ovoid symbol. Squared boxes denote unitary actions
performed after that exchange.\\ Related results have been
recently derived by other authors \cite{benni1}. Reznik et al
studied the deterministic remote implementation of a class of
operations whose complete specification is split between the
remote partners. For instance, single qubit transformations of the
form $U=e^{i \alpha_A \sigma_{\hat{n}_B}}$, where the rotation
angle $\alpha_A$ is controlled by Alice while Bob selects the
direction $\hat{n}_{B}$. In this case, the most economical
protocol for the remote implementation of $U$ consumes 1 ebit and
requires the symmetrical exchange of two classical bits. Reznik
and Groisman have also analyzed the probabilistic remote
implementation of controlled operations using partially entangled
states as a resource \cite{benni2}. The potential use of this type
of protocols for secret sharing schemes was discussed by
Gea-Banacloche and C-P Yang \cite{gea}.

\subsection{Efficient application of multiple instances of an
unknown unitary operation}
%%%%%%%%%%%%%%%%%%%%%%%%%%%%%%%%%%%%%%%%%%%%%%%%%%%%%%%%%%%%%%%%%%%Telecloning %%%%%%%%%%%%%%%%%%%%%%%%
In this subsection we are going to present a novel result
concerning the generalization of quantum remote control. So far we
have considered the question of a single remote application of an
unknown unitary operation on an individual quantum state. One
might consider whether the remote application of $U$ on two
identical copies of a quantum state can be carried out with fewer
resources than two full ebits and two classical bits of
communication in each direction (ie twice the resource of a single
application). In the following we will show that indeed, as long
as the two copies of the quantum state $|\psi\rangle$ are both
held by Bob a resource reduction can be achieved. In other words,
assuming Alice holds a machine that implements
$U=e^{i\theta\sigma_z}$ with an unknown $\theta$ and given that
Bob holds the state $|\psi\rangle^{\otimes 2}$ we would like to
provide an entangled state shared between Alice and Bob which,
when supplemented by local operations and classical communication,
allows for Bob to hold, at the end of the protocol, the state
$(U|\psi\rangle)^{\otimes 2}$. In the following we will provide a
protocol that requires an entangled state with $\log 3$ ebits of
entanglement. This is clearly less than 2 ebits and therefore
represents a resource reduction over the trivial protocol.

The entangled resource is the well-known tele-cloning state
\cite{Murao JPV 99}
\begin{equation}
    |\psi_{tele}\rangle = \frac{1}{\sqrt{3}}\left(|00\rangle_A|00\rangle_B +
    \frac{|01\rangle_A+|10\rangle_A}{\sqrt{2}}\frac{|01\rangle_B+|10\rangle_B}{\sqrt{2}}
    + |11\rangle_A|11\rangle_B\right)
\end{equation}
For the following it will be convenient to use the following
abbreviations
\begin{eqnarray*}
    |\tilde{0}\rangle &=& |00\rangle\\
    |\tilde{1}\rangle &=& \frac{|01\rangle+|10\rangle}{\sqrt{2}}\\
    |\tilde{2}\rangle &=& |11\rangle
\end{eqnarray*}
Then the total state including $|\psi\rangle^{\otimes 2}$ with
$|\psi\rangle = \alpha|0\rangle+\beta|1\rangle$ is given by
\begin{equation}
    |\psi_{tele}\rangle|\psi\rangle^{\otimes 2} =
    \frac{1}{\sqrt{3}}(|\tilde{0}\rangle_A|\tilde{0}\rangle_B
    + |\tilde{1}\rangle_A|\tilde{1}\rangle_B + |\tilde{2}\rangle_A
    |\tilde{2}\rangle_B)(\alpha^2|\tilde{0}\rangle+\sqrt{2}\alpha\beta|\tilde{1}\rangle
    +\beta^2|\tilde{2}\rangle)
\end{equation}
let us further define the three unitary transformations
\begin{eqnarray}
    \id &=& |\tilde{0}\rangle\langle \tilde{0}|+ |\tilde{1}\rangle\langle \tilde{1}|
    + |\tilde{2}\rangle\langle \tilde{2}|\\
    T &=& |\tilde{0}\rangle\langle \tilde{1}|+ |\tilde{1}\rangle\langle \tilde{2}|
    + |\tilde{2}\rangle\langle \tilde{0}|\\
    T^2 &=& |\tilde{0}\rangle\langle \tilde{2}|+ |\tilde{1}\rangle\langle \tilde{0}|
    + |\tilde{2}\rangle\langle \tilde{1}|
\end{eqnarray}
and the corresponding controlled operation
\begin{equation}
    U_T = \sum_{k=0}^{2} |\tilde{k}\rangle\langle \tilde{k}|\otimes T^k
\end{equation}
We furthermore define the unitary operator
\begin{equation}
    {\cal F} = \frac{1}{\sqrt{3}}\left(\begin{array}{ccc}
    1   &      1     & 1\\
    1   & e^{2i\pi/3} & e^{4i\pi/3}\\
    1   & e^{4i\pi/3} & e^{8i\pi/3}
    \end{array}\right)
\end{equation}
It is now straightforward to verify that the following protocol
will implement $U$ on the two copies of $|\psi\rangle$. (i) Apply
$U_T$ between the first (second) copy of $|\psi\rangle$ as control
and the first (second) qubit of Bob's part of the telecloning
state as target. (ii) Measure Bob's qubits that are part of the
telecloning state. If he finds $|\tilde{k}\rangle$, then apply
operation $T^k$ on Alice's qubits. (iii) Now she applies $U$ to
both of her qubits. (iv) Alice applies ${\cal F}$ and subsequently
performs a measurement in the
$\{|\tilde{0}\rangle,|\tilde{1}\rangle,|\tilde{2}\rangle\}$ basis.
(v) If Alice finds $|\tilde{k}\rangle$, then Bob needs to apply
the operation ${\cal F}^k$ on his qubits.\\ The outcome of this
procedure is the state $(U|\psi\rangle)^{\otimes 2}$ on Bob's
side.

The procedure requires the transmission of $\log 3$ classical bits
in both directions and furthermore requires the telecloning state
as a resource which contains $\log 3$ ebit of entanglement.

Note however, that in this protocol is is strictly necessary that
Bob holds both copies of the state $|\psi\rangle$ and possesses
the ability to carry out joint operations on both of them. These
joint operations would require require further entangled resources
if the two particles on Bob's side are distant. It is therefore
natural to consider the question of whether one can remotely apply
the unitary operation $U$ that Alice possesses onto two identical
copies of the state $|\psi\rangle$ that are held by Bob and
Charles. If one admits a $50\%$ success rate then this is indeed
possible employing a straightforward generalization of the
protocol for remote application of $U=e^{i\theta\sigma_z}$
presented earlier in this section. It is an interesting open
question whether there is a protocol that achieves $100\%$ success
probability {\em and} requires less than a shared ebit between
Alice and Bob and another shared ebit between Alice and Charles.

\section{Recent experimental results}
Recently, the quantum remote control protocol for arbitrary single
qubit rotations, shown in Figure 3, was implemented experimentally
in a linear optics setup \cite{Xiang LG 05}. For that,
polarization entangled states generated from spontaneous
parametric down conversion (SPDC) where locally manipulated to
generate a three qubit state involving polarization and path
degrees of freedom. In the following we will describe the basic
elements of this experiment as well as the results that have been
obtained with this set-up.

We begin with a brief description of the experimental choice of
the physical qubit, the gates implementing the local operations
and the entangled states employed.
\begin{itemize}

\item {\em Qubits:} For photons, both horizontal and vertical
polarization states $\left\{ \left\vert H\right\rangle ,\left\vert
V\right\rangle \right\} $ as well as up and down paths $\left\{
\left\vert u\right\rangle ,\left\vert d\right\rangle \right\} $
can represent the logic states $\{\left\vert 0\right\rangle
,\left\vert 1\right\rangle \}$ of qubits. We will refer to the
resulting encoding as polarization or path qubits respectively.

\item {\em Quantum gates:} For a polarization qubit, arbitrary
unitary rotation can be performed by using half-wave plate(HWP)
and quarter-wave plate(QWP)\cite{JKMW}. The controlled-NOT gate
between polarization qubit (control) and path qubit (target) of
the same photon can be implemented by a polarization beam
splitter(PBS),
\begin{eqnarray}
\left\vert H\right\rangle \left\vert u\right\rangle (\left\vert
H\right\rangle \left\vert d\right\rangle )&\rightarrow& \left\vert
H\right\rangle \left\vert u^{\prime }\right\rangle (\left\vert
H\right\rangle \left\vert d^{\prime }\right\rangle ),\\ \nonumber
\left\vert V\right\rangle \left\vert u\right\rangle (\left\vert
V\right\rangle \left\vert d\right\rangle )&\rightarrow& \left\vert
V\right\rangle \left\vert d^{\prime }\right\rangle (\left\vert
V\right\rangle \left\vert u^{\prime }\right\rangle
),\end{eqnarray} with suitable definition of the incoming and
outgoing modes.

\item {\em Entangled states:} Bi-photon entangled states involving
either polarization or path qubits can be generated via a SPDC
process \cite{K}. Here the initial three qubit state of the
protocol represented in Fig. 3 will involves the polarization
degree of freedom of Alice's qubit and the polarization and path
degrees of freedom of Bob's particle, as explained in detail
below.
\end{itemize}

\begin{figure}[htb]
\epsfxsize=8.7cm
\begin{center}
\epsffile{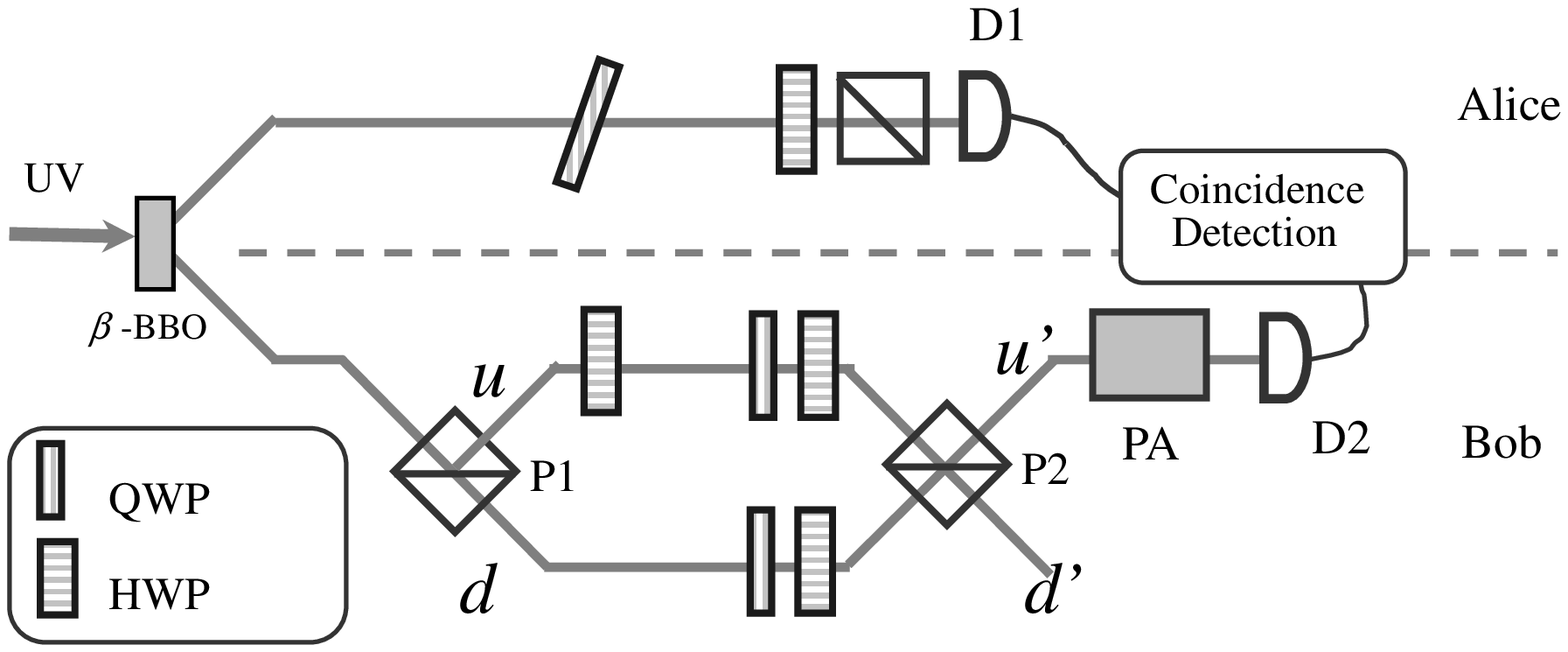}
\end{center}
\caption{\label{setup} Schematic representation of the
experimental setup to perform a remote rotation on a single photon
\cite{Xiang LG 05}. \textit{P1}, \textit{P2} denote polarization
beam splitters; HWP are half wave plates; QWP are quarter wave
plates; PA is a polarization analyzer; $D_{1},D_{2}$ represent
single photon detectors.}
\end{figure}

The concrete experimental setup that was implemented in
\cite{Xiang LG 05} is showed schematically in Fig. \ref{setup}. A
mode-locked Ti:Sapphire pulsed laser (with the pulse width less
than 200 fs, a repetition rate of about 82MHz and a central
wavelength of 780.0nm) is frequency-doubled to produce the pumping
source for a SPDC process. A BBO crystal of 1mm thickness, cut for
type-II phase matching, is used as a down converter. Non-collinear
degenerated SPDC generates two photons, \textit{A} and\textit{\
B}, in the
polarization-entangled state $$\left\vert \Psi ^{+}\right\rangle _{AB}=\frac{1%
}{\sqrt{2}}(\left\vert H\right\rangle _{A}\left\vert
V\right\rangle _{B}+ \left\vert V\right\rangle _{A}\left\vert
H\right\rangle _{B})$$\cite{K}. Bob employs the PBS (denoted
\textit{P1}) to split photon \textit{B} in two paths $\left\{
\left\vert u\right\rangle ,\left\vert d\right\rangle \right\} $
and a HWP \textit{H1} at $45^{\circ }$ as a $\sigma _{x}$ gate\ is
used to flip the polarization in path \textit{u}. Hence the
initial polarization entanglement between the two distributed
photons is converted into polarization-path entanglement,
\begin{equation}
\left\vert \Psi ^{+}\right\rangle
_{123}=\frac{1}{\sqrt{2}}(\left\vert H\right\rangle _{1}\left\vert
u\right\rangle _{2}+\left\vert V\right\rangle _{1}\left\vert
d\right\rangle _{2})\left\vert H\right\rangle _{3}, \label{ES}
\end{equation}%
where we have relabelled Alice's photon with the index $1$ and the
the indices $2$ and $3$ refer to the path and polarization degree
of freedom of photon B. The polarization state
of qubit $3$ can be prepared in an arbitrary state $%
\left\vert \psi \right\rangle _{3}=\alpha \left\vert
H\right\rangle _{3}+\beta \left\vert V\right\rangle _{3}$ with
identical sets of waveplates, $\{H_{u},Q_{u}\}$ and
$\{H_{d},Q_{d}\}$, in each path \cite{JKMW}. The global state can
therefore be initialized to be of the general form
\begin{equation}
\left\vert \Phi ^{+}\right\rangle _{12}\left\vert \psi
\right\rangle _{3}=\frac{1}{\sqrt{2}}(\left\vert H\right\rangle
_{1}\left\vert u\right\rangle _{2}+\left\vert V\right\rangle
_{1}\left\vert d\right\rangle _{2})(\alpha \left\vert
H\right\rangle _{3}+\beta \left\vert V\right\rangle
_{3}).\end{equation} The three step protocol for the remote
rotation of Bob's polarization qubit is performed as follows:\\

\textit{i}) \textit{Encoding (Operation $G_1$)}: The paths
\textit{u} and \textit{d} of photon \textit{B} provide the input
for a second PBS (denoted by \textit{P2}) to perform a
\textit{CNOT} operation, where the polarization acts as the
control qubit and the path represents the target qubit. In the
experiment we have ensured that the optical path lengths of
\textit{u} and \textit{p} are equal to avoid the accumulation of a
relative phase factor between the two terms in eq. (\ref{ES}). The
$\sigma _{z}$ measurement on qubit \textit{2} is implemented by
reading out the path information of photon \textit{B}. If
\textit{B} is in path \textit{u}', $\left\vert \psi \right\rangle
_{B}$ is encoded into $\left\vert \psi \right\rangle _{AB}=\alpha
\left\vert H\right\rangle _{A}\left\vert H\right\rangle _{B}+\beta
\left\vert V\right\rangle _{A}\left\vert V\right\rangle _{B}$. If
\textit{B} is in path \textit{d}', the two photons will be in
$\left\vert \psi ^{\prime }\right\rangle _{13}=\alpha \left\vert
V\right\rangle _{1}\left\vert H\right\rangle _{3}+\beta \left\vert
H\right\rangle _{1}\left\vert V\right\rangle _{3}$, which can be
transformed into $\left\vert \Psi \right\rangle _{13}$ by another
HWP at $45^{\circ }$ acting on photon \textit{A}. Here we omit the
later case without loss of generality.
%The polarization state of photon
%\textit{B} is encoded in $\left\{ \left\vert H\right\rangle
%_{1}\left\vert H\right\rangle _{3},\left\vert V\right\rangle
%_{1}\left\vert V\right\rangle _{3}\right\} $\cite{PJF}.\\

\textit{ii}) \textit{Remote operation}: The operation $U_{com}$
can be performed
by a pair of QWP at $45^{\circ }$ with a HWP at $\frac{\varphi }{2}%
-45^{\circ }$ between them. Such device has been used to verify
the geometric phase of classical light and photons\cite{HR,BDM}.
For a single qubit operation, any additional global phase is
trivial, so $U_{com}$ can be replaced by $e^{i\varphi /2}U_{com}$,
which can be realised by one zero-order waveplate at $0^{\circ }$
tilted in a suitable angle (see \cite {KWWAE} for similar
application). Here we chose $\varphi =60^{\circ }$ and $120^{\circ }$ by a tilted
QWP \textit{Q1}.\\

\textit{iii}) \textit{Decoding and Verification (Operation
$G_2$)}: Alice performs her measurement in the rotated basis
$\{\left\vert D\right\rangle _{1}=\frac{1}{\sqrt{2}}\left(
\left\vert H\right\rangle
_{1}+\left\vert V\right\rangle _{1}\right) ,\left\vert C\right\rangle _{1}=%
\frac{1}{\sqrt{2}}\left( \left\vert H\right\rangle _{1}-\left\vert
V\right\rangle _{1}\right) \}$ using a polarizer. Photon
\textit{A} is detected by a single photon detector (SPCM-AQR-14 by
EG\&G). Photon \textit{B} will be collapsed into $\left\vert \psi
^{\prime }\right\rangle _{3}=U_{com}\left\vert \psi \right\rangle
_{3}$ for result $\left\vert +\right\rangle _{1}$, and $\left\vert
\psi ^{\prime \prime }\right\rangle
_{3}=U_{com}\sigma _{z}\left\vert \psi \right\rangle _{3}$ for result $%
\left\vert -\right\rangle _{1}$. The latter can be converted into $%
\left\vert \psi ^{\prime }\right\rangle _{B}$ by a HWP at
$0^{\circ }$, i.e. a $\sigma _{z}$ rotation. The polarization
state of photon \textit{B} is reconstructed by quantum state
tomography using a polarization analyzer and a detector
\cite{JKMW}. The measurements on \textit{A} and \textit{B} are
collected via coincidence counts with a window time of 5ns.\\

{\em Results:} \vspace*{0.5cm}

\begin{figure}[htb]
\epsfxsize=10.7cm
\begin{center}
\epsffile{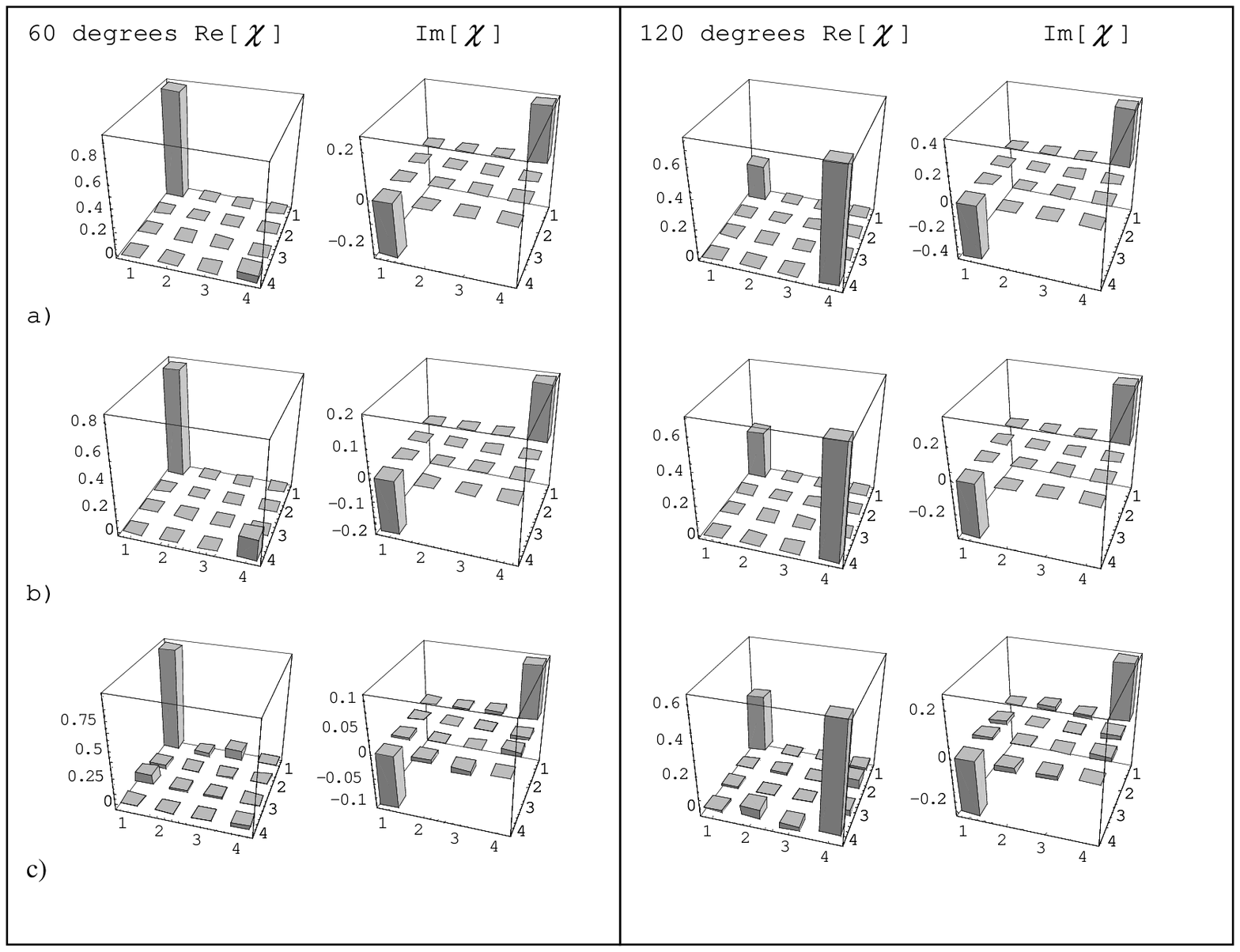}
\end{center}
\caption{\label{results} Quantum process tomography for the remote
$\sigma_{z}$ rotation of $60^{\circ }$ and $120^{\circ }$. We have
represented the theoretical values of the $\chi$ matrices
corresponding to (a) ideal rotation $\chi_i$ and (b) a dephased
rotation $\chi_d$  as well as c) the measured rotation $\chi _{e}$
performed with the set up depicted in Fig. 4. The real parts of
the matrix elements of $\chi $ are represented in the chart
figures on the left while the imaginary ones are represented on
the right.}
\end{figure}
%%%%%%%%%%%%%%%%%%%%%%%%%%%%%%%%%%%%%%%%%%%%%%%%%%%%%%%%%%%%%%%%%%%%%% New text Jian Li %%%%%%%%%%%%%%%%%%%%%%%%%%%%%%%%%%%%%%%%%%%%%%
The effect of a general quantum operation on a single qubit can
represented by a trace-preserving completely positive (CP) map,
i.e. for an arbitrary input state $\rho$, the output one would be
of the form  $\rho^{\prime}= \varepsilon (\rho )=\sum_{mn}\chi
_{mn}E_{m}\rho E_{n}^{\dagger }$,\, ($\{E_{m}\}=\{I,\sigma
_{x},\sigma _{y},\sigma _{z}\}$) , where $\chi $ is a positive
Hermitian matrix. In the actual experiment, two main sources of
phase decoherence were identified (for a more detailed discussion
see \cite{Xiang LG 05}). One is caused by the bi-refringency of
BBO, which induces the partial time-separation between the
wave-packets of two polarizations. The other one is the mismatch
of spatial modes in the PBS \textit{P2}. The CP map representing
the dephasing operation can
be written as ${\Large \varepsilon }_{d}%
{\Large =}\{\sqrt{\frac{1+p\eta }{2}}U_{com},\sqrt{\frac{1-p\eta }{2}}%
U_{com}\sigma _{z}\}$ where $p$ is the visibility of the entangled
state obtained from SPDC and $\eta$ is the visibility of the
interferometer formed by the PBSs $P1$ and $P2$. These parameters
can be measured independently. Within this formalism, the final
state after the action of the net quantum operation, including
dephasing, representing the remote rotation protocol is given by
\begin{equation}
\rho _{d}=\varepsilon _{d}(\ket{\psi} \bra{\psi})=\left(
\begin{array}{cc}
\alpha \alpha ^{\ast } & p\eta \alpha \beta ^{\ast }e^{-i\varphi } \\
p\eta \alpha ^{\ast }\beta e^{i\varphi } & \beta \beta ^{\ast }%
\end{array}%
\right) .
\end{equation}
Here, to completely characterize the remote operation $\varepsilon
_{e}$ in our experiment, four states $\{\left\vert H\right\rangle
,\left\vert V\right\rangle ,\left\vert D\right\rangle ,\left\vert
R\right\rangle = \frac{1}{\sqrt{2}}\left( \left\vert
H\right\rangle -i\left\vert V\right\rangle \right) \}$ are used as
an input for the initial state of Bob's polarization qubit.
%For each input state, the output state is measured using
%quantum state tomography.
With the four output density matrices, the $\chi $ obtained in the
process can then be reconstructed using techniques for quantum
process tomography \cite{ABJ}. For the $\sigma_{z}$-rotation
$U_{com}=\cos\phi/2+i \sin\phi/2\sigma_{z}$, $\chi_{i}$ has four
non-zero elements, $\chi_{11}=(1+\cos\phi)/2$,
$\chi_{44}=(1-\cos\phi)/2$, $\chi^{\ast}_{14}=\chi_{41}=i
\sin\phi/2$. The dephasing only changes the value of the four
non-zero elements, $\chi_{11}=(1+p \,\eta \,cos\phi)/2$,
$\chi_{44}=(1-p\,\eta\, cos\phi)/2$, $\chi^{\ast}_{14}=\chi_{41}=i
p\,\eta \,sin\phi$, while leaving the other 12 zero elements
unaffected. The matrices are shown in Fig. \ref{results} for the
case of an ideal rotation $\chi_{i}$, a dephased rotation $\chi
_{d}$, and the effective operation $\chi _{e}$ obtained in our
experiment, where the left six histograms are the real parts for
the matrices for $60^{\circ}$ and $120^{\circ}$ rotation through
$z$ axis, and the right for the imaginary ones, where the values
of the parameter $\varepsilon _{d}$ was measured to be $p\approx
0.85$ and $\eta \approx 0.92$. New non-zero elements are found in
the effective $\chi_{e}$. This is introduced by the imperfection
of the polarization beam-splitter. The comparison of the
experimental operation $ \varepsilon_{e}$ with the ideal rotation
$\varepsilon _{i}$ is determined by evaluating the average
fidelity with pure input states uniformly distributed over the
Bloch sphere $$ \overline{F}[\varepsilon_{e},U_{com} ]=\int d\psi
F[\varepsilon _{e}(\ket{\psi}, U_{com}\ket{\psi} \bra{\psi}
U_{com}^{\dagger}]$$ where
$$F[\rho ,\rho ^{\prime}]=(Tr[\sqrt{\sqrt{\rho ^{\prime }}\rho
\sqrt{\rho ^{\prime }}}])^2$$
 is the output state fidelity
\cite{BOSBJ,N}.
The measured $\chi$ yields $\overline{F}_{60}=0.96$ and $%
\overline{F}_{120}=0.86$. Although only a rotation commuting with
$\sigma_{z} $ ($z$-rotation) is operated in our experiment, the
same protocol can be implemented for operations anti-commuting
with $\sigma_{z}$ ($x$-rotation) with another HWP at $45^{\circ} $
at the output port, which acts as a $\sigma_{z} $ operation.  The
above scheme can also be generalised to implementing controlled
operations which commute or anti-commute with $\sigma_{z}$. For
example, the control-phase gate commutes with $I\otimes \sigma
_{z}$ , $\sigma _{z}\otimes I$ and $\sigma _{z}\otimes \sigma
_{z}$. An experiment along these lines, where a non-local CNOT was
implemented using linear optics, was reported in \cite{Huang}.

Summarizing, we have implemented a remote rotation by $120^{\circ
}$ about the $z$ axis on a photonic qubit using shared
entanglement and local operations and without performing the
rotation directly on the target photons. The whole process was
characterized using quantum process tomography and the results
agree with the theoretical predictions. The scheme can be
generalized to implement remotely any operation belonging to the
two classes that allow for protocols different from bidirectional
state teleportation \cite{angles}.

\section{Conclusions}

The linearity of quantum mechanics imposes severe constraints on
the type of protocols that can be implemented by quantum
mechanical means. When combining these constraints with those of
locality, one arrives at interesting no-go theorems concerning
quantum protocols between spatially separated parties. In this
work we have investigated such questions both theoretically and
experimentally. In particular, we have considered the possibility
of the remote implementation of unitary transformations. If Alice
holds a tool that allows for the implementation of a unitary
transformation we consider the question whether entanglement and
classical communication is sufficient to allow this unitary
transformation be applied to an unknown state of a particle held
by Bob. We have considered several scenarios and, employing
linearity, the non-increase of entanglement under local operations
and classical communication (LOCC) as well as the no-signalling
condition, several no-go theorems were derived and discussed. We
have also identified situations with restricted sets of operations
in which the remote application of unitaries can be achieved in a
non-trivial way, that is, avoiding state teleportation.
%These results
%illuminate the constraints imposed on us by fundamental properties
%of quantum mechanics.
Despite their theoretical nature, the results of this work have
led to an effort towards the implementation of such protocols and
we have described the successful implementation of one of the
theoretical protocols that has been developed in this paper.

It is hoped that these results illuminate further how fundamental
properties of quantum mechanics impose constraints via linearity,
locality and no-signalling but also facilitate new opportunities
in the form of generalized measurements and entanglement
supplemented by classical communication. The use of multipartite
entanglement, as discussed in section III.a, opens up a new front
and further applications for quantum communication are
foreseeable.

 \acknowledgements The authors
would like to thank Claire Bedrock for her patience and her
willingness to 'enjoy the challenge' of dealing with a very late
submission. This work was supported by the IRC on Quantum
Information of the EPSRC (GR/S82176/0), The EU via the Integrated
Project QAP, the Thematic Network QUPRODIS (IST-2002-38877), the
Leverhulme Trust (F/07058/U), the Chinese National Fundamental
Research Program, the NSF of China and the Innovation Funds from
the Chinese Academy of Sciences.

%\end{multicols}

\end{document}